\documentclass[12pt,a4paper,twoside]{article}
\usepackage{graphics}
\usepackage{delarray}
\usepackage{rotating}
\newlength{\figwidth}
\newlength{\figheight}
\title{Contact Interactions, Large Extra Dimensions and Leptoquarks at THERA}
\author{Aleksander Filip \.Zarnecki \\
{\small\it Institute of Experimental Physics, Warsaw University,} \\
{\small\it Ho\.za 69, 00-681 Warszawa, Poland} \\
{\small\it E-mail: zarnecki@fuw.edu.pl}
} 
\begin{document} 

\maketitle 

\begin{abstract}

The sensitivity of THERA to different models of ``new physics''
has been studied, both for the contact interaction approximation and
for the resonance production.
For contact interaction models conserving parity, 
scales up to about 18 TeV can be 
explored at THERA, extending considerably beyond the existing bounds.
Significant improvement of existing limits is also expected for models
with large extra dimensions.
Effective Plank mass scales up to about 2.8 TeV can be probed.
THERA will be the best machine to study leptoquark properties,
for leptoquark masses up to about 1 TeV.
It will be sensitive to the leptoquark Yukawa couplings down to
$\lambda_{LQ} \sim 10^{-2}$.

\end{abstract}

\thispagestyle{empty}

%
%---------------------------------------------------------------------------
%

\section{Introduction}
\label{sec:afz:intro}

Search for "new physics" has always been one of the most 
important subjects in the field of particle physics. 
Possibility of discovering new particles, new interactions and/or
other new phenomena is always considered as a main
argument for building new, more powerful colliders.
Collected in this paper are results concerning possible ``new physics''
searches at THERA, prepared as a part of the dedicated THERA physics study.
The collider project and running options are briefly summarized in
Sect.~\ref{sec:afz:tera}. Models considered in this paper are introduced 
in Sect.~\ref{sec:afz:model}.
The method used for the analysis has been developed for 
the global analysis of the existing data \cite{epj:c11:539,epj:c17:695}
and is briefly described in Sect.~\ref{sec:afz:method}.
Some of the results discussed in Sect.~\ref{sec:afz:results}
have been already presented in \cite{hep-ph-0006335}.

%---------------------------------------------------------------------------
%
\section{THERA collider}
\label{sec:afz:tera}

THERA has been proposed as the next generation $ep$ collider,
a straightforward extension of the TESLA project.
It would bring high-energy electrons or positrons 
from the linear accelerator 
into collisions with high-energy protons from the existing HERA
collider.
Using both arms of TESLA center-of-mass energies of up
to 1.6 TeV could be obtained at THERA, exceeding the energy
range currently accessible at HERA by up to a factor of five.

Following THERA running scenarios are considered in this paper:
\begin{itemize}

\item   
For nominal electron beam energy of $E_{e}$=250 GeV and
proton beam energy of $E_{p}$=1000 GeV 
integrated luminosity of about 40 pb$^{-1}$ is expected in a year.
Results presented for this option
assume integrated luminosity of 100  pb$^{-1}$
for $e^{-}p$ and/or  100  pb$^{-1}$ for $e^{+}p$ collisions.
It will be referred to as {\bf\sc Thera-250}.

\item
Using both arms of TESLA, electron beam energy can be increased
to  $E_{e}$=500 GeV. 
Assumed integrated luminosity for this scenario is also 100  pb$^{-1}$
for $e^{-}p$ and/or  100  pb$^{-1}$ for $e^{+}p$ collisions.
It will be referred to as {\bf\sc Thera-500}.

\item 
With TESLA machine upgraded in power, electron energies as high
as  $E_{e}$=800 GeV are possible for THERA operation.
In this case proton beam energy is lowered to $E_{p}$=800 GeV,
to provide maximum luminosity.
Results presented for this option
assume integrated luminosity of 200  pb$^{-1}$
for $e^{-}p$ and/or  200  pb$^{-1}$ for $e^{+}p$ collisions.
It will be referred to as {\bf\sc Thera-800}.

\end{itemize}

%

%
%---------------------------------------------------------------------------
\section{Models of new physics}
\label{sec:afz:model}

\subsection{Contact Interactions}
\label{sec:afz:ci}

Four-fermion contact interactions are an effective theory, which 
allows us to describe, in the most general way, possible low energy 
effects  coming from "new physics" at much higher energy scales. 
%
% This includes the possible existence of second generation heavy weak bosons, 
% leptoquarks as well as electron and quark compositeness \cite{cidef,hera91-haberl}.
%
As very strong limits have been already placed on both scalar 
and tensor contact interactions \cite{hera91-haberl}, only vector 
contact interactions are considered in this analysis.
The influence of the vector contact interactions on the $ep$ NC DIS 
cross-section can be described as an additional term 
in the tree level $eq \rightarrow eq$ scattering amplitude~\cite{pr:d57:391,hep-ph-9810277}:
\begin{eqnarray}
M^{e_i q_j \rightarrow e_i q_j}(t) & = & 
- \frac{4 \pi \alpha_{em} e_{q}}{t} \; + \;
  \frac{4 \pi \alpha_{em}}{sin^{2}\theta_{W} \cdot cos^{2}\theta_{W} } 
\cdot \frac{g^{e}_{i} g^{q}_{j}}{t - M_{Z}^{2}}
  \; + \; \eta^{eq}_{ij} \label{eq:afz:mt}
\end{eqnarray}
where $t = -Q^{2}$ is the Mandelstam variable describing the
four-momentum transfer between the electron and the quark, $e_{q}$ is 
the electric charge of the quark in units of the elementary charge,
the subscripts $i$ and $j$ label the chiralities of the initial 
lepton and quark respectively ($i,j=L,R$), $g^{e}_{i}$ and $g^{q}_{j}$ 
are electroweak couplings of the electron and the quark,
and $\eta^{eq}_{ij}$ are the contact interaction couplings.

In the most general case, vector contact interactions are described by
4 independent couplings for every quark flavor. 
As $ep$ scattering is sensitive predominantly 
to electron-up and  electron-down quark couplings,
8 independent couplings should be considered.
However, it is not be possible, in one experiment,
to put significant constraints on all of these couplings 
simultaneously, without additional assumptions.
Therefor, only one-parameter models, assuming fixed relations between 
the separate couplings, are considered in this paper.
Relations between couplings assumed for different models 
are presented in Tab.~\ref{tab:afz:cimod1} and \ref{tab:afz:cimod2}.
Listed in Tab.~\ref{tab:afz:cimod1} are models with defined 
coupling chirality. Models in Tab.~\ref{tab:afz:cimod2}
fulfill the relation
\begin{eqnarray}
\eta^{eq}_{LL} + \eta^{eq}_{LR} - \eta^{eq}_{RL} - \eta^{eq}_{RR}
                    & = &  0  \nonumber
\end{eqnarray}
which is imposed to conserve parity, and to avoid strong limits coming 
from atomic parity violation measurements.
In the presented contact interaction analysis it is also assumed that 
all up type and down type quarks have the same contact interaction couplings:
\begin{eqnarray}
\eta^{eu}_{ij} & = \eta^{ec}_{ij} & = \eta^{et}_{ij} \nonumber \\ 
% label{eq:afz:qu} \\
\eta^{ed}_{ij} & = \eta^{es}_{ij} & = \eta^{eb}_{ij} \nonumber
\end{eqnarray}
Coupling $\eta$ can be related to the effective 
mass scale of contact interactions $\Lambda$: 
\begin{eqnarray}
\eta & = & \pm \frac{g_{CI}^{2}}{\Lambda^{2}}   \nonumber
\end{eqnarray}
where the coupling strength of new interactions is by convention
set to $g_{CI} = \sqrt{4\pi}$.

\begin{table}[p]
  \begin{center}
   \begin{tabular}{crrrrrrrr}
      \hline\hline\hline\noalign{\smallskip}
  Model & 
$\eta^{ed}_{LL}$ & $\eta^{ed}_{LR}$ & $\eta^{ed}_{RL}$ & $\eta^{ed}_{RR}$ & 
$\eta^{eu}_{LL}$ & $\eta^{eu}_{LR}$ & $\eta^{eu}_{RL}$ & $\eta^{eu}_{RR}$ \\ 
\hline\hline\hline\noalign{\smallskip}
$q_{LL}$ &   $+\eta$ &   &   &   & $+\eta$ &  &  &   \\
$q_{LR}$ &   &   $+\eta$ &   &   &  & $+\eta$ &  &   \\
$q_{RL}$ &   &   &   $+\eta$ &   &  &  & $+\eta$ &   \\
$q_{RR}$ &   &   &   &   $+\eta$ &  &  &  &  $+\eta$ \\
\hline\noalign{\smallskip}
$d_{LL}$ &   $+\eta$ &   &   &   &  &  &  &   \\
$d_{LR}$ &   &   $+\eta$ &   &   &  &  &  &   \\
$d_{RL}$ &   &   &   $+\eta$ &   &  &  &  &   \\
$d_{RR}$ &   &   &   &   $+\eta$ &  &  &  &   \\
\hline\noalign{\smallskip}
$u_{LL}$ &   &   &   &   & $+\eta$ &  &  &   \\
$u_{LR}$ &   &   &   &   &  & $+\eta$ &  &   \\
$u_{RL}$ &   &   &   &   &  &  & $+\eta$ &   \\
$u_{RR}$ &   &   &   &   &  &  &  &  $+\eta$ \\
\hline\hline\hline\noalign{\smallskip}
    \end{tabular}
  \end{center}
  \caption{Relations between couplings for contact interaction models 
           with defined coupling chirality considered in this paper.}
  \label{tab:afz:cimod1}
\end{table}

\begin{table}[p]
  \begin{center}
   \begin{tabular}{crrrrrrrr}
      \hline\hline\hline\noalign{\smallskip}
  Model & 
$\eta^{ed}_{LL}$ & $\eta^{ed}_{LR}$ & $\eta^{ed}_{RL}$ & $\eta^{ed}_{RR}$ & 
$\eta^{eu}_{LL}$ & $\eta^{eu}_{LR}$ & $\eta^{eu}_{RL}$ & $\eta^{eu}_{RR}$ \\ 
\hline\hline\hline\noalign{\smallskip}
VV & $+\eta$& $+\eta$& $+\eta$& $+\eta$& $+\eta$& $+\eta$& $+\eta$& $+\eta$\\
AA & $+\eta$& $-\eta$& $-\eta$& $+\eta$& $+\eta$& $-\eta$& $-\eta$& $+\eta$\\
VA & $+\eta$& $-\eta$& $+\eta$& $-\eta$& $+\eta$& $-\eta$& $+\eta$& $-\eta$\\
\hline\noalign{\smallskip}
X1 & $+\eta$ & $-\eta$ &  &  & $+\eta$ & $-\eta$ &  &  \\
X2 & $+\eta$ &  & $+\eta$ &  & $+\eta$ &  & $+\eta$ &  \\
X3 & $+\eta$ &  &  & $+\eta$ & $+\eta$ &  &  & $+\eta$ \\
X4 &  & $+\eta$ & $+\eta$ &  &  & $+\eta$ & $+\eta$ &  \\
X5 &  & $+\eta$ &  & $+\eta$ &  & $+\eta$ &  & $+\eta$ \\
X6 &  &  & $+\eta$ & $-\eta$ &  &  & $+\eta$ & $-\eta$ \\ 
\hline\noalign{\smallskip}
U1 &   &   &   &   & $+\eta$ & $-\eta$ &  &  \\
U2 &   &   &   &   & $+\eta$ &  & $+\eta$ &  \\
U3 &   &   &   &   & $+\eta$ &  &  & $+\eta$ \\
U4 &   &   &   &   &  & $+\eta$ & $+\eta$ &  \\
U5 &   &   &   &   &  & $+\eta$ &  & $+\eta$ \\
U6 &   &   &   &   &  &  & $+\eta$ & $-\eta$ \\ 
\hline\hline\hline\noalign{\smallskip}
    \end{tabular}
  \end{center}
  \caption{Relations between couplings for the parity conserving
         contact interaction models considered in this paper.}
  \label{tab:afz:cimod2}
\end{table}

%---------------------------------------------------------------------------

\subsection{Large Extra Dimensions}
\label{sec:afz:ed}

Model proposed by Arkani-Hamed, Dimopoulos and Dvali 
\cite{pl:b429:263,pr:d59:086004} assumes the
space-time is $4+n$ dimensional. Standard Model particles, including strong and
electroweak bosons are confined to 4 dimensions, but the gravity can
propagate in the extra dimensions as well.
With very large extra dimensions, the effective Plank scale $M_{S}$ can be
of the order of TeV.
The graviton, after summing the effects of its excitations 
in the extra dimensions, couples to the Standard Model particles 
with an effective strength of $1/M_{S}$.
At high energies,
gravitation interaction can become comparable in strength to electroweak
interactions.
Virtual graviton exchange contribution to $eq \rightarrow eq$ scattering
can be described by an effective contact interactions.
Contribution to the scattering amplitude (\ref{eq:afz:mt}),
equivalent to the cross-section formula given in \cite{pl:b460:383}, 
can be written as:
\begin{eqnarray}
\eta^{eq}_{LL} & = \; \eta^{eq}_{RR}  &  = \; 
  - \frac{ \pi \lambda}{ 2  M_{S}^{4}}  \left( 4t + s \right) \nonumber \\ 
%  \label{eq:afz:cied}
\eta^{eq}_{LR} & = \; \eta^{eq}_{RL}  &  = \; 
- \frac{ \pi \lambda }{ 2  M_{S}^{4}} \left( 4t + 3s \right) \nonumber
\end{eqnarray}
where $t$ and $s$ are the Mandelstam variables describing 
electron-quark scattering.
By convention the coupling strength is set to $\lambda = \pm 1$.

%---------------------------------------------------------------------------

\subsection{Leptoquarks}
\label{sec:afz:lq}

In this paper a general classification of leptoquark states 
proposed by Buchm\"uller, R\"uckl and Wyler \cite{pl:b191:442} will be used.
The Buchm\"uller-R\"uckl-Wyler (BRW) model is based on 
the assumption that new interactions should respect the 
$SU(3)_{C} \times SU(2)_{L} \times U(1)_{Y}$ symmetry of
the Standard Model.
In addition leptoquark couplings are assumed to be family diagonal
(to avoid FCNC processes) and to conserve lepton and baryon numbers
(to avoid rapid proton decay).
Taking into account very strong bounds from rare decays \cite{pl:b177:377}
it is also assumed that leptoquarks couple either to left-
or to right-handed leptons.
With all these assumptions there are 14 possible states 
(isospin singlets or multiplets) of scalar and vector leptoquarks.
Table~\ref{tab:afz:aachen} lists these states according to 
the so-called Aachen notation \cite{zfp:c46:679}.
An S(V) denotes a scalar(vector) leptoquark and the subscript
denotes the weak isospin.
When the leptoquark can couple to both right- and left-handed
leptons, an additional superscript indicates  the lepton chirality.
A tilde is introduced to differentiate between leptoquarks
with different hypercharge.
Listed in Tab.~\ref{tab:afz:aachen} are the leptoquark fermion
number F, electric charge Q, and the branching ratio to an electron-quark
pair (or electron-antiquark pair), $\beta$.
The leptoquark branching fractions are predicted by the BRW model 
and are either 1, $\frac{1}{2}$ or 0.
For a given electron-quark branching ratio $\beta$, the branching ratio 
to the neutrino-quark is by definition $(1-\beta)$. 
Also included in Tab.~\ref{tab:afz:aachen} are the flavours and chiralities of 
the lepton-quark pairs coupling to a given leptoquark type.
In three cases the squark flavours (in supersymmetric theories with 
broken R-parity) with corresponding  couplings
are also indicated.
Present analysis takes into account only leptoquarks which couple
to the first-generation leptons ($e$, $\nu_{e}$) and first-generation 
quarks ($u$, $d$).
It is also assumed that one of the leptoquark 
types gives the dominant contribution, as compared with other leptoquark
states  and that the interference between different leptoquark 
states can be neglected.
Using this simplifying assumption, different leptoquark types 
can be considered separately.
Finally, it is assumed that different leptoquark states within 
isospin doublets and triplets have the same mass.
 
\begin{table}[p]
 \begin{center}
   \begin{tabular}{lcccccc}
      \hline\hline\hline\noalign{\smallskip}
Model & Fermion & Charge & $BR(LQ \rightarrow e^{\pm}q)$ & 
        \multicolumn{2}{c}{ } & Squark \\
      & number F &   Q   & $\beta$  &  \multicolumn{2}{c}{Coupling} & type \\
\hline\hline\hline\noalign{\smallskip}
$S_{\circ}^L$ &  2  &  $-1/3$  &  1/2  &   $e_{L}u$ & $\nu d$  & $\tilde{d_R}$ \\
\hline\noalign{\smallskip}
$S_{\circ}^R$ &  2  &  $-1/3$  &  1  &  $e_{R}u$ &  &   \\
\hline\noalign{\smallskip}
$\tilde{S}_{\circ}$  &  2  &  $-4/3$   &  1  &  $e_{R}d$ & &    \\
\hline\noalign{\smallskip}
$S_{1/2}^L$   &  0  &  $-5/3$   &  1  &  $e_{L} \bar{u}$ & &  \\
              &     &  $-2/3$   &  0  &  & $\nu \bar{u}$  &   \\
\hline\noalign{\smallskip}
$S_{1/2}^R$   &  0  &  $-5/3$  &  1 &  $e_{R} \bar{u}$  &  &  \\
              &     &  $-2/3$  &  1 &  $e_{R} \bar{d}$  &  &  \\
\hline\noalign{\smallskip}
$\tilde{S}_{1/2}$ &  0 &  $-2/3$  &  1  &  $e_{L} \bar{d}$  &  & $\overline{\tilde{u}_{L}}$  \\
                  &    &  $+1/3$  &  0  &     & $\nu \bar{d}$  & $\overline{\tilde{d}_{L}}$  \\
\hline\noalign{\smallskip}
$S_{1}$       &  2  & $-4/3$  &  1  & $e_{L}d$ & & \\
              &     & $-1/3$  &  1/2 & $e_{L}u$ &  $\nu d$  & \\
              &     & $+2/3$  &  0   &  &  $\nu u$  &  \\
\hline\hline\hline\noalign{\smallskip}
$V_{\circ}^L$ &  0  &  $-2/3$  &  1/2  &   $e_{L}\bar{d}$ & $\nu \bar{u}$ & \\
\hline\noalign{\smallskip}
$V_{\circ}^R$ &  0  &  $-2/3$  &  1  &  $e_{R}\bar{d}$ & & \\
\hline\noalign{\smallskip}
$\tilde{V}_{\circ}$  &  0  &  $-5/3$   &  1  &  $e_{R}\bar{u}$  & &  \\
\hline\noalign{\smallskip}
$V_{1/2}^L$   &  2  &  $-4/3$   &  1  &  $e_{L} d$  & & \\
              &     &  $-1/3$   &  0  &  & $\nu d$  &   \\
\hline\noalign{\smallskip}
$V_{1/2}^R$   &  2  &  $-4/3$  &  1 &  $e_{R} d$  &  &   \\
              &     &  $-1/3$  &  1 &  $e_{R} u$  &  &  \\
\hline\noalign{\smallskip}
$\tilde{V}_{1/2}$ &  2 &  $-1/3$  &  1  &  $e_{L} u$ &  &  \\
                 &     &  $+2/3$   &  0  &  & $\nu u$  &   \\
\hline\noalign{\smallskip}
$V_{1}$       &  0  & $-5/3$  &  1  & $e_{L} \bar{u}$ & \\
              &     & $-2/3$  &  1/2 & $e_{L}\bar{d}$ & $\nu \bar{u}$ & \\
              &     & $+1/3$  &  0   &   &  $\nu \bar{d}$  &  \\
\hline\hline\hline\noalign{\smallskip}
    \end{tabular}
   \end{center}
  \caption{A general classification of leptoquark states 
 in the Buchm\"uller-R\"uckl-Wyler model. 
 Listed are the leptoquark fermion number, F, 
 electric charge, Q (in units of elementary charge), 
the branching ratio to electron-quark (or electron-antiquark), $\beta$
 and the flavours of the coupled lepton-quark pairs. 
Also shown are possible squark assignments to the leptoquark states
 in  the minimal supersymmetric theories with broken R-parity.}
  \label{tab:afz:aachen}
\end{table}

In the limit of heavy leptoquark masses ($M_{LQ} \gg \sqrt{s}$)
the effect of leptoquark production or exchange
is equivalent to a vector type $eeqq$ contact interaction.
Contribution to the $eq\rightarrow eq$ scattering amplitude (\ref{eq:afz:mt})
does not depend on the process kinematics and can be written as
\begin{eqnarray}
\eta^{eq}_{ij} & = & 
a^{eq}_{ij} \cdot \left( \frac{\lambda_{LQ}}{M_{LQ}} \right)^{2} \;\; ,
\nonumber
% \label{eq:afz:cieta}
\end{eqnarray}
where $M_{LQ}$ is the leptoquark mass, $\lambda_{LQ}$ 
the leptoquark-electron-quark Yukawa coupling 
and  the coefficients $a^{eq}_{ij}$ are 
given in Tab.~\ref{tab:afz:lqci} \cite{zfp:c74:595}.

\begin{table}[p]
 \begin{center}
   \begin{tabular}{lrrrrrrrr}
      \hline\hline\hline\noalign{\smallskip}
  Model & 
$a^{ed}_{LL}$ & $a^{ed}_{LR}$ & $a^{ed}_{RL}$ & $a^{ed}_{RR}$ & 
$a^{eu}_{LL}$ & $a^{eu}_{LR}$ & $a^{eu}_{RL}$ & $a^{eu}_{RR}$ \\ 
\noalign{\smallskip}
\hline\hline\hline\noalign{\smallskip}
$S_{\circ}^L$ &    &    &    &    & +$\frac{1}{2}$   &    &    &    \\
$S_{\circ}^R$ &    &    &    &    &    &    &    &  +$\frac{1}{2}$     \\
$\tilde{S}_{\circ}$ 
              &    &    &    &  +$\frac{1}{2}$     &    &    &    &    \\
$S_{1/2}^L$   &    &    &    &    &    &  $-\frac{1}{2}$     &    &    \\
$S_{1/2}^R$   &    &    &  $-\frac{1}{2}$ &   &   &   & $-\frac{1}{2}$ &    \\
$\tilde{S}_{1/2}$ 
              &    &  $-\frac{1}{2}$  &    &    &    &    &    &    \\
$S_{1}$       &  +1  &    &    &    &  +$\frac{1}{2}$   &    &    &    \\
\noalign{\smallskip}
\hline\hline\hline\noalign{\smallskip}
$V_{\circ}^L$ &  $-1$  &    &    &    &    &    &    &    \\
$V_{\circ}^R$ &    &    &    &  $-1$  &    &    &    &    \\
$\tilde{V}_{\circ}$ 
              &    &    &    &    &    &    &    &  $-1$   \\
$V_{1/2}^L$   &    &  +1  &    &    &    &    &    &    \\
$V_{1/2}^R$   &    &    &  +1  &    &    &    &  +1   &    \\
$\tilde{V}_{1/2}$ 
              &    &    &    &    &    &  +1  &    &    \\
$V_{1}$       & $-1$   &    &    &    &  $-2$  &    &    &    \\
\noalign{\smallskip}
\hline\hline\hline\noalign{\smallskip}
    \end{tabular}
   \end{center}
  \caption{Coefficients $a^{eq}_{ij}$ defining the effective 
           contact interaction couplings 
 $\eta^{eq}_{ij}=a^{eq}_{ij}\cdot \frac{\lambda_{LQ}^2}{M_{LQ}^2}$ 
 for different models of scalar (upper part of the table)
 and vector (lower part) leptoquarks.           
 Empty places in the table correspond to $a^{eq}_{ij}=0$.}
  \label{tab:afz:lqci}
\end{table}

\begin{figure}[p]
\centerline{\resizebox{!}{9cm}{%
  \includegraphics{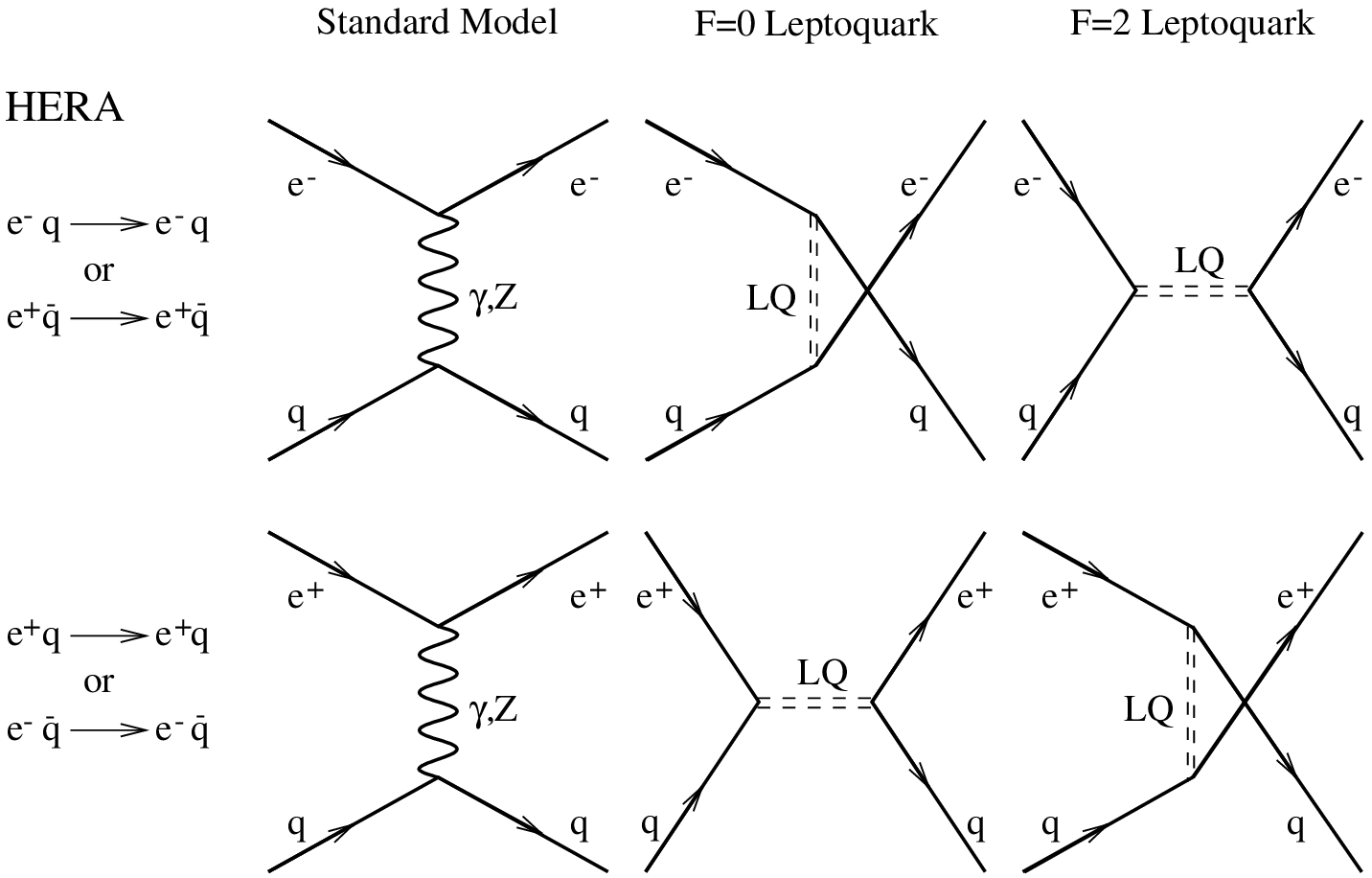}
}}
  \caption{Diagrams describing leading order Standard Model processes
           and leptoquark contributions coming from F=0 and F=2 leptoquarks,
           for NC $e^{\pm}p$ DIS at THERA.}
  \label{fig:afz:lqdiag}
\end{figure}

For leptoquark masses comparable with the available $ep$ center-of-mass energy
$u$-channel leptoquark exchange process and the $s$-channel leptoquark 
production have to be considered separately. 
Corresponding diagrams for F=0 and F=2 leptoquarks are shown in
Fig.~\ref{fig:afz:lqdiag}.
The leptoquark contribution to the scattering amplitude 
can be now described by the following formulae:
\begin{itemize}
\item
for $u$-channel leptoquark exchange 
( $F$=0 leptoquark in $e^{-}q$ or $e^{+}\bar{q}$ scattering, 
  or $|F|$=2 leptoquark in $e^{+}q$ or $e^{-}\bar{q}$ scattering)
\begin{eqnarray}
\eta^{eq}_{ij}(s,u) & = & \frac{a^{eq}_{ij} \cdot \lambda_{LQ}^{2}}
                            { M_{LQ}^{2} - u }  \;\; , \nonumber
\end{eqnarray}

\item
for $s$-channel leptoquark production 
($F$=0 leptoquark in $e^{+}q$ or $e^{-}\bar{q}$ scattering, 
 or $|F|$=2 leptoquark in $e^{-}q$ or $e^{+}\bar{q}$ scattering)
\begin{eqnarray}
\eta^{eq}_{ij}(s,u) & = & \frac{a^{eq}_{ij} \cdot \lambda_{LQ}^{2}}
{M_{LQ}^{2} - s - i s \frac{\Gamma_{LQ}}{M_{LQ}}} \;\; , \nonumber
\end{eqnarray}
where $\Gamma_{LQ}$ is the total leptoquark width. 
The partial decay width for every decay channel is given by the formula:
\begin{eqnarray}
\Gamma_{LQ} & = & \frac{ \lambda_{LQ}^{2} M_{LQ} }
                       { 8 \pi ( J + 2 ) }  \;\; , \nonumber
\end{eqnarray}
where $J$ is the leptoquark spin.
\end{itemize}

For leptoquark masses smaller than the available $ep$ center-of-mass energy
direct production of single leptoquarks can be considered.
In the narrow-width approximation, the cross-section for single 
$F=2$ leptoquark production in electron-proton scattering 
(via the electron-quark fusion) is given by:
\begin{eqnarray}
\sigma^{ep\rightarrow LQ \; X}(M_{LQ},\lambda_{LQ}) & = & 
(J+1) \cdot \frac{\pi \lambda_{LQ}^{2}}{4 M_{LQ}^{2}} 
        \cdot x_{LQ} q(x_{LQ},M_{LQ}^{2}) \label{eq:afz:dirlq}
\end{eqnarray}
where $q(x,Q^{2})$ is the quark momentum distribution
in the proton and $x_{LQ} =\frac{M_{LQ}^{2}}{s}$.

%---------------------------------------------------------------------------
%
\section{Analysis method}
\label{sec:afz:method} 

The analysis method used has been described in details in the recently 
published papers \cite{epj:c11:539,epj:c17:695}.
For all models considered, limits on the model parameters 
can be extracted from the measured $Q^{2}$ distribution
of NC DIS events at THERA. 
The leading-order doubly-differential cross-section for electron-proton 
NC DIS ($e^{-}p \rightarrow e^{-} X$) can be written as 
\begin{eqnarray}
\frac{d^{2}\sigma^{LO}}{dx dQ^{2}} & 
= \displaystyle{\frac{1}{16\pi} \sum_{q}} & 
q(x,Q^{2}) \left\{ |M^{eq}_{LL}|^{2} + |M^{eq}_{RR}|^{2} 
%           \right.     \; \; \; \; + \nonumber \\  & & \left. 
  (1-y)^{2} \left[  |M^{eq}_{LR}|^{2} + |M^{eq}_{RL}|^{2} \right] 
    \right\} \; +  \nonumber \\
 & & \bar{q}(x,Q^{2})\left\{ |M^{eq}_{LR}|^{2} + |M^{eq}_{RL}|^{2} 
%                     \right. \; \; \; \; + \nonumber \\ & & \left. 
  (1-y)^{2} \left[  |M^{eq}_{LL}|^{2} + |M^{eq}_{RR}|^{2} \right] 
             \right\} \;\; ,  \nonumber
\end{eqnarray}
where $x$ is the Bjorken variable, describing the fraction of 
the proton momentum  carried by the struck quark (antiquark), 
$q(x,Q^{2})$ and $\bar{q}(x,Q^{2})$ are
the quark and antiquark momentum distribution functions in the proton and
$M^{eq}_{ij}$ are the scattering amplitudes of
(\ref{eq:afz:mt}), which can include contributions from 
``new physics''.

The cross-section integrated over the $x$ and $Q^{2}$ range 
of an experimental $Q^{2}$ bin from $Q^{2}_{min}$ to $Q^{2}_{max}$ is
\begin{eqnarray}
\sigma^{LO} & = &  \int\limits_{Q^{2}_{min}}^{Q^{2}_{max}} dQ^{2} 
\int\limits_{\frac{Q^{2}}{s \cdot y_{max}}}^{1} dx \;  
\frac{d^{2}\sigma^{LO}}{dx dQ^{2}} \label{eq:afz:intdis}
\end{eqnarray}
where $y_{max}$ is an upper limit on the reconstructed Bjorken variable $y$,  
$y=\frac{Q^{2}}{x\;s}$. In the presented analysis this limit is set
to $y_{max}$=0.95.
Eq.~\ref{eq:afz:intdis} is used to calculate numbers of expected 
events in  $Q^{2}$ bins.
Expected limits on model parameters,
from high-$Q^{2}$ NC $e^{\pm}p$ DIS at THERA,
are calculated assuming that
no deviations from the Standard Model predictions will be observed.

For every value of the model parameter $\eta$ 
($\pm 4\pi/\Lambda^{2}$ for the contact interaction models,
$\pm 1/M_{S}^{4}$ for large extra dimensions, $(\lambda_{LQ}/M_{LQ})^{2}$ for
leptoquark models in the high-mass limit\footnote{
For leptoquark masses comparable with the available center-of-mass energy,
two parameter probability function ${\cal P}(\lambda_{LQ},M_{LQ})$ is
considered and limits on $\lambda_{LQ}$ are calculated as a function of
the leptoquark mass $M_{LQ}$.
}) the probability function
describing the agreement between the model and the data is calculated:
\begin{eqnarray}
{\cal P}(\eta) & \sim & 
\prod_{i} P_{i}(\eta) \nonumber
%  \label{eq:afz:prob}
\end{eqnarray}
The product runs over all $Q^{2}$ bins $i$ (separately for
$e^{-}p$ and $e^{+}p$ data). 
The probability $P_{i}$
described by the Poisson distribution
\begin{eqnarray}
P_{i}(\eta) & \sim & 
\frac{ n(\eta)^{N} \cdot exp( -n(\eta)) }
     { N ! }  , \label{eq:afz:poisson} 
\end{eqnarray}
where $N$ and $n(\eta)$ are the measured and expected 
number of events in a given bin.
This formula properly takes into account statistical errors
in the measured event distributions. 
The systematic errors in the Standard Model expectations are
assumed to be correlated to 100\%
between different $Q^{2}$ bins, and increase from 1\% at
$Q^{2}$=1000 GeV$^{2}$ to 5\% at $Q^{2}$=100000 GeV$^{2}$.
The method used to include systematic errors, as well as the 
migration corrections resulting from the assumed $Q^{2}$ measurement 
resolution of 5\%  are discussed in detail in \cite{epj:c11:539}.

To constraint leptoquark Yukawa coupling values, 
for leptoquark masses smaller than the available $ep$ center-of-mass energy, 
direct production of single leptoquarks is also considered,
as described by (\ref{eq:afz:dirlq}).
Only the leptoquark signal in the electron-jet decay channel is
taken into account.
Expected signal from single leptoquark production,
for given leptoquark mass $M_{LQ}$ and Yukawa coupling $\lambda_{LQ}$, 
is compared with the observed number of events from the Standard Model 
background (NC DIS) in the $\pm$5\% mass window.
The background is suppressed by applying a cut on 
the Bjorken variable $y$, which is optimized for every leptoquark 
type as a function of the leptoquark mass.
After the $y$ cut is imposed, probability function  
${\cal P}(\lambda_{LQ},M_{LQ})$
is calculated from the Poisson distribution (\ref{eq:afz:poisson}).
As ${\cal P}$ is not a probability distribution, it does
not satisfy any normalization condition.
Instead it is convenient to rescale the probability function in such a
way that for the Standard Model it has the value of 1:
\begin{eqnarray}
{\cal P}(\eta = 0)  & = &  1. \nonumber
% \label{eq:afz:pmax} 
\end{eqnarray}

Using the probability function ${\cal P}(\eta)$ 
limits on the model parameters are calculated.
Rejected are all models (parameter values) which result in
\begin{eqnarray}
{\cal P}(\eta)  & < &  0.05  \label{eq:afz:plim} 
\end{eqnarray}
This is taken as the definition of the 95\% confidence level (CL)
exclusion limit.\footnote{
For Gaussian shape of the probability function, condition 
(\ref{eq:afz:plim}) corresponds to $\pm 2.45 \sigma$ limit.
Mass scale limits presented in this paper
would increase by 10 to 15\%, if the definition 
more commonly used in the literature is used:
${\cal P}(\eta) = 0.147$ corresponding to $\pm 1.96 \sigma$.
However, this definition assumes the Gaussian shape 
of the probability function, which is not always the case.
Therefor definition (\ref{eq:afz:plim}) is used as more ``conservative''.
Same limit setting method has been used in the global analysis of
existing data \cite{epj:c17:695,hep-ph-0006196}.
} 
Exclusion limits presented in this paper 
are lower limits in case of mass scales $\Lambda$ or
$M_{S}$, leptoquark mass $M_{LQ}$ or   $M_{LQ}/\lambda_{LQ}$,  and
upper limits in case of $\lambda_{LQ}$.
For leptoquark masses smaller than the available center-of-mass energy,
both indirect (from $\frac{d\sigma}{dQ^{2}}$) and direct 
$\lambda_{LQ}$ limits are calculated, and the stronger one is presented.

%
%---------------------------------------------------------------------------
%
\section{Results}
\label{sec:afz:results}

\addtolength{\tabcolsep}{0.3mm}
\begin{sidewaystable}[p]
 \begin{center}
   \begin{tabular}{lrrcrrrrrrcrrrrrrcrrrrrr}
      \hline\hline\hline\noalign{\smallskip}
  &  \multicolumn{2}{c}{ Current} &
  &  \multicolumn{20}{c}{ Expected 95\% CL exclusion limits [TeV]} \\ 
                                 \cline{5-24}\noalign{\smallskip}
 Model   &  \multicolumn{2}{c}{limits} & 
 \multicolumn{6}{c}{\sc Thera-250} & & \multicolumn{6}{c}{\sc Thera-500} & &
   \multicolumn{6}{c}{\sc Thera-800} \\
 \cline{5-10} \cline{12-17}  \cline{19-24}\noalign{\smallskip}
 & \multicolumn{2}{c}{[TeV]}  &
 & \multicolumn{2}{c}{$e^{-}p$}  
 & \multicolumn{2}{c}{$e^{+}p$}  
 & \multicolumn{2}{c}{$e^{\pm}p$} &
 & \multicolumn{2}{c}{$e^{-}p$}  
 & \multicolumn{2}{c}{$e^{+}p$}  
 & \multicolumn{2}{c}{$e^{\pm}p$} &
 & \multicolumn{2}{c}{$e^{-}p$}  
 & \multicolumn{2}{c}{$e^{+}p$}  
 & \multicolumn{2}{c}{$e^{\pm}p$} \\
% \cline{4-21} 
\noalign{\smallskip}
 & $\Lambda^{-}$ & $\Lambda^{+}$ &
 & $\Lambda^{-}$ & $\Lambda^{+}$ 
 & $\Lambda^{-}$ & $\Lambda^{+}$ 
 & $\Lambda^{-}$ & $\Lambda^{+}$ &
 & $\Lambda^{-}$ & $\Lambda^{+}$ 
 & $\Lambda^{-}$ & $\Lambda^{+}$ 
 & $\Lambda^{-}$ & $\Lambda^{+}$ &
 & $\Lambda^{-}$ & $\Lambda^{+}$ 
 & $\Lambda^{-}$ & $\Lambda^{+}$ 
 & $\Lambda^{-}$ & $\Lambda^{+}$ \\
\noalign{\smallskip}
\hline\hline\hline\noalign{\smallskip}
%
% Results from files:
% tera_250_ele_ci.tex tera_250_pos_ci.tex tera_250_comb_ci.tex tera_500_ele_ci.tex tera_500_pos_ci.tex tera_500_comb_ci.tex tera_800_ele_ci.tex tera_800_pos_ci.tex tera_800_comb_ci.tex
%
$d_{LL}$     &   23.3  &    8.4  &
    & 5.6 & 4.7 & 5.0 & 4.2 & 6.3 & 5.5 &
    & 7.3 & 6.0 & 6.5 & 5.6 & 8.1 & 7.1 &
    & 8.8 & 7.4 & 7.8 & 6.8 & 9.8 & 8.7 \\
$d_{LR}$     &   19.6  &    7.5  &
   & 3.2 & 2.9 & 4.8 & 3.9 & 4.9 & 4.0 &
    & 4.1 & 3.7 & 6.0 & 5.0 & 6.1 & 5.1 &
    & 4.9 & 4.4 & 7.3 & 6.0 & 7.5 & 6.1 \\
$d_{RL}$     &    7.4  &   20.4  &
  & 2.7 & 3.3 & 3.9 & 4.8 & 3.9 & 4.9 &
    & 3.5 & 4.3 & 5.0 & 6.0 & 5.0 & 6.2 &
    & 4.2 & 5.1 & 6.0 & 7.4 & 6.0 & 7.5 \\
$d_{RR}$     &    8.8  &   17.5  &
  & 4.6 & 3.5 & 4.0 & 3.0 & 4.9 & 3.6 &
    & 5.8 & 4.5 & 5.2 & 3.8 & 6.3 & 4.7 &
    & 7.0 & 5.3 & 6.2 & 4.5 & 7.6 & 5.6 \\
\hline\noalign{\smallskip}
$u_{LL}$     &   14.7  &   11.3  &
 & 7.9 & 8.3 & 5.6 & 5.9 & 8.5 & 8.8 &
    & 9.8 & 10.3 & 7.0 & 7.6 & 10.5 & 10.9 &
    & 12.2 & 12.7 & 8.7 & 9.2 & 13.0 & 13.5 \\
$u_{LR}$     &   16.9  &    7.8  &
   & 3.1 & 3.8 & 4.9 & 7.1 & 5.1 & 7.3 &
    & 3.9 & 4.9 & 6.0 & 8.7 & 6.2 & 8.9 &
    & 4.6 & 5.8 & 7.4 & 10.9 & 7.6 & 11.0 \\
$u_{RL}$     &    7.1  &   18.3   &
  & 3.2 & 3.4 & 4.7 & 6.2 & 4.9 & 6.3 &
    & 4.0 & 4.5 & 6.0 & 7.7 & 6.2 & 7.8 &
    & 4.7 & 5.3 & 7.3 & 9.4 & 7.4 & 9.6 \\
$u_{RR}$     &    7.0  &   21.2  &
 & 6.3 & 7.3 & 4.4 & 5.3 & 6.8 & 7.8 &
    & 7.6 & 9.1 & 5.6 & 6.8 & 8.4 & 9.7 &
    & 9.6 & 11.3 & 6.8 & 8.3 & 10.4 & 11.9 \\
\hline\noalign{\smallskip}
$q_{LL}$     &   26.1  &   10.9   &
 & 4.2 & 7.2 & 3.5 & 4.8 & 4.3 & 7.3 &
    & 5.3 & 8.8 & 4.6 & 6.1 & 5.5 & 9.0 &
    & 6.4 & 10.9 & 5.4 & 7.4 & 6.6 & 11.2 \\
$q_{LR}$     &   25.8  &   10.6  &
  & 3.5 & 4.0 & 4.9 & 7.1 & 5.0 & 7.2 &
    & 4.5 & 5.2 & 6.3 & 8.7 & 6.4 & 8.8 &
    & 5.3 & 6.2 & 7.6 & 10.7 & 7.7 & 10.9 \\
$q_{RL}$     &   10.2  &   27.3   &
 & 3.4 & 4.1 & 5.1 & 7.0 & 5.2 & 7.2 &
    & 4.4 & 5.4 & 6.5 & 8.7 & 6.6 & 8.9 &
    & 5.2 & 6.4 & 7.8 & 10.7 & 8.0 & 10.9 \\
$q_{RR}$     &   10.3  &   27.1   &
 & 4.4 & 7.0 & 3.5 & 5.0 & 4.5 & 7.3 &
 & 5.6 & 8.7 & 4.5 & 6.3 & 5.8 & 9.0 &
 & 6.8 & 10.7 & 5.4 & 7.7 & 7.0 & 11.1 \\
\noalign{\smallskip}
\hline\hline\hline\noalign{\smallskip}
    \end{tabular}
   \end{center}
  \caption{95\% CL exclusion limits on the contact interaction
         mass scales  $\Lambda^{-}$ and $\Lambda^{+}$ (for
         negative and positive coupling signs respectively)
         expected from the
         measurement of high-$Q^{2}$ NC DIS cross-sections 
         at  THERA, for different running scenarios, as indicated
         in the table. Limits from the global analysis of existing
          data \protect\cite{epj:c11:539,hep-ph-0006196}
          are included for comparison.
           }
  \label{tab:afz:cires1}
\end{sidewaystable}
\addtolength{\tabcolsep}{-0.3mm}

\addtolength{\tabcolsep}{-0.1mm}
\begin{sidewaystable}[p]
 \begin{center}
   \begin{tabular}{lrrcrrrrrrcrrrrrrcrrrrrr}
      \hline\hline\hline\noalign{\smallskip}
  &  \multicolumn{2}{c}{ Current} &
  &  \multicolumn{20}{c}{ Expected 95\% CL exclusion limits [TeV]} \\ 
                                 \cline{5-24}\noalign{\smallskip}
 Model   &  \multicolumn{2}{c}{limits} & 
 \multicolumn{6}{c}{\sc Thera-250} & & \multicolumn{6}{c}{\sc Thera-500} & &
   \multicolumn{6}{c}{\sc Thera-800} \\
 \cline{5-10} \cline{12-17}  \cline{19-24}\noalign{\smallskip}
 & \multicolumn{2}{c}{[TeV]}  &
 & \multicolumn{2}{c}{$e^{-}p$}  
 & \multicolumn{2}{c}{$e^{+}p$}  
 & \multicolumn{2}{c}{$e^{\pm}p$} &
 & \multicolumn{2}{c}{$e^{-}p$}  
 & \multicolumn{2}{c}{$e^{+}p$}  
 & \multicolumn{2}{c}{$e^{\pm}p$} &
 & \multicolumn{2}{c}{$e^{-}p$}  
 & \multicolumn{2}{c}{$e^{+}p$}  
 & \multicolumn{2}{c}{$e^{\pm}p$} \\
% \cline{4-21} 
\noalign{\smallskip}
 & $\Lambda^{-}$ & $\Lambda^{+}$ &
 & $\Lambda^{-}$ & $\Lambda^{+}$ 
 & $\Lambda^{-}$ & $\Lambda^{+}$ 
 & $\Lambda^{-}$ & $\Lambda^{+}$ &
 & $\Lambda^{-}$ & $\Lambda^{+}$ 
 & $\Lambda^{-}$ & $\Lambda^{+}$ 
 & $\Lambda^{-}$ & $\Lambda^{+}$ &
 & $\Lambda^{-}$ & $\Lambda^{+}$ 
 & $\Lambda^{-}$ & $\Lambda^{+}$ 
 & $\Lambda^{-}$ & $\Lambda^{+}$ \\
\noalign{\smallskip}
\hline\hline\hline\noalign{\smallskip}
%
% Results from files:
% tera_250_ele_ci.tex tera_250_pos_ci.tex tera_250_comb_ci.tex tera_500_ele_ci.tex tera_500_pos_ci.tex tera_500_comb_ci.tex tera_800_ele_ci.tex tera_800_pos_ci.tex tera_800_comb_ci.tex
%
$VV $     &    8.3  &   14.5 &
  &  7.6 & 9.9 & 7.2 & 10.3 & 9.8 & 11.9 &
  & 8.7 & 12.3 & 8.5 & 12.5 & 11.3 & 14.6 &
  & 11.3 & 15.2 & 10.6 & 15.6 & 14.4 & 18.1 \\
$AA $     &   11.2  &   10.8 &
  & 4.7 & 9.2 & 8.4 & 6.2 & 9.3 & 9.3 &
  & 6.2 & 11.2 & 10.4 & 8.1 & 11.1 & 11.4 &
  & 7.3 & 14.0 & 12.8 & 9.7 & 13.9 & 14.1 \\
$VA $     &    5.8  &    6.3 &
  & 6.3 & 6.7 & 7.3 & 7.2 & 7.7 & 7.7 &
  & 8.2 & 8.5 & 9.3 & 9.1 & 9.8 & 9.9 &
  & 9.7 & 10.3 & 11.3 & 11.0 & 11.8 & 11.9 \\
\hline\noalign{\smallskip}
$X1 $     &    8.5  &    8.6   &
  & 4.2 & 7.2 & 6.8 & 5.4 & 7.2 & 7.3 &
  & 5.6 & 8.8 & 8.5 & 7.0 & 8.7 & 9.1 &
  & 6.5 & 10.9 & 10.4 & 8.4 & 10.8 & 11.2 \\
$X2 $     &    6.7  &   10.8 &
  & 4.5 & 7.6 & 5.2 & 7.8 & 6.1 & 9.0 &
  & 5.7 & 9.5 & 6.7 & 9.6 & 7.4 & 11.1 &
  & 6.9 & 11.6 & 8.1 & 11.9 & 9.2 & 13.6 \\
$X3 $     &    8.8  &   12.0  &
  & 7.0 & 9.4 & 4.0 & 6.2 & 7.4 & 9.7 &
  & 7.9 & 11.5 & 5.2 & 7.9 & 8.2 & 11.9 &
  & 10.5 & 14.2 & 6.1 & 9.6 & 11.0 & 14.7 \\
$X4 $     &    6.2  &   10.0   &
  & 4.0 & 5.0 & 5.9 & 9.1 & 6.1 & 9.2 &
  & 5.2 & 6.6 & 7.5 & 11.1 & 7.7 & 11.3 &
  & 6.1 & 7.8 & 9.1 & 13.7 & 9.2 & 13.9 \\
$X5 $     &    5.6  &    9.1  &
  & 4.5 & 7.4 & 5.1 & 7.9 & 5.8 & 8.9 &
  & 5.8 & 9.2 & 6.6 & 9.8 & 7.2 & 11.0 &
  & 6.9 & 11.3 & 7.9 & 12.0 & 8.8 & 13.6 \\
$X6 $     &    6.8  &    5.4 &
  & 6.9 & 4.4 & 5.5 & 6.7 & 7.1 & 6.9 &
  & 8.6 & 5.7 & 7.1 & 8.4 & 9.0 & 8.6 &
  & 10.5 & 6.8 & 8.6 & 10.3 & 10.9 & 10.5 \\
\hline\noalign{\smallskip}
$U1 $     &    6.3  &   13.0 &
  & 7.1 & 8.1 & 6.2 & 5.1 & 8.1 & 8.1 &
  & 8.3 & 9.9 & 7.7 & 6.7 & 9.7 & 10.1 &
  & 10.7 & 12.3 & 9.5 & 8.0 & 12.2 & 12.5 \\
$U2 $     &    7.3  &   15.6 &
  & 8.1 & 8.5 & 6.6 & 7.9 & 9.1 & 9.7 &
  & 9.9 & 10.6 & 8.1 & 9.9 & 11.1 & 12.1 &
  & 12.3 & 13.1 & 10.0 & 12.1 & 13.8 & 14.9 \\
$U3 $     &    8.9  &   19.8  &
  & 10.8 & 10.8 & 7.6 & 7.7 & 11.5 & 11.4 &
  & 13.3 & 13.3 & 9.6 & 9.9 & 14.2 & 14.2 &
  & 16.5 & 16.4 & 11.8 & 11.9 & 17.5 & 17.5 \\
$U4 $     &    5.2  &    8.4  &
  & 3.8 & 4.5 & 6.2 & 8.7 & 6.5 & 8.9 &
  & 4.7 & 5.9 & 7.3 & 10.6 & 7.6 & 10.8 &
  & 5.5 & 7.0 & 9.0 & 13.2 & 9.3 & 13.3 \\
$U5 $     &    6.9  &   14.8 &
  & 6.9 & 7.8 & 7.3 & 8.4 & 8.7 & 9.6 &
  & 8.4 & 9.8 & 8.7 & 10.3 & 10.6 & 11.9 &
  & 10.4 & 12.0 & 10.8 & 12.8 & 13.2 & 14.7 \\
$U6 $     &   11.9  &    5.8   &
  & 7.3 & 5.3 & 5.5 & 5.7 & 7.6 & 6.8 &
  & 9.1 & 6.3 & 7.1 & 7.2 & 9.5 & 8.3 &
  & 11.2 & 8.1 & 8.6 & 8.8 & 11.6 & 10.4 \\
\hline\hline\hline\noalign{\smallskip}
$ED $ &   &  0.94   &
  & 1.41 & 1.74 & 1.81 & 1.52 & 1.88 & 1.83 &
  & 1.84 & 2.29 & 2.36 & 1.96 & 2.44 & 2.38 &
  & 2.18 & 2.70 & 2.81 & 2.34 & 2.90 & 2.83 \\
\noalign{\smallskip}
\hline\hline\hline\noalign{\smallskip}
    \end{tabular}
   \end{center}
  \caption{95\% CL exclusion limits on the contact interaction
         mass scales  $\Lambda^{-}$ and $\Lambda^{+}$, and
         on the the effective Plank scales $M^{-}_{S}$ and  $M^{+}_{S}$ 
         in the large extra dimensions (ED) model (for
         negative and positive coupling signs respectively)
         expected from the
         measurement of high-$Q^{2}$ NC DIS cross-sections 
         at  THERA, for different running scenarios, as indicated
         in the table. Limits from the global analysis of existing
         data \protect\cite{epj:c11:539,hep-ph-0006196,pl:b460:383}
          are included for comparison.
           }
  \label{tab:afz:cires2}
\end{sidewaystable}
\addtolength{\tabcolsep}{0.1mm}

95\% CL exclusion limits on the contact interaction mass scales
$\Lambda^{-}$ and $\Lambda^{+}$ (for negative and positive coupling signs) 
expected from the measurement of high-$Q^{2}$ NC DIS cross-sections 
at THERA, are presented in Tab.~\ref{tab:afz:cires1} and  \ref{tab:afz:cires2}.
Results presented are the mean values from 1000 MC experiments.
Poisson fluctuations in the observed numbers of events can result in
the statistical fluctuations in the limit values of the order of 10-20\%.  
Current limits from the global contact interaction analysis 
\cite{hep-ph-0006196}\footnote{
Numerical limit values presented in this paper differ slightly from
limits presented in  \cite{hep-ph-0006196}. They have been recalculated using
data on $eeqq$ interactions only. Data from neutrino scattering experiments and
from charged current processes, which can be included in the analysis
when assuming $SU(2)_{L} \times U(1)_{Y}$ symmetry of new interactions,
were not used. This is because some of the considered models violate $SU(2)$
invariance.
}
and from the global analysis of existing data 
in the large extra dimensions (ED) model \cite{pl:b460:383}
are included for comparison.

For contact interaction models violating parity (models with defined 
coupling chirality; Tab.~\ref{tab:afz:cires1}),
current limits from global analysis are already of the order
of 10-20 TeV. 
This is mainly due to very strong constraints from the atomic
parity violation (APV)  measurements in cesium 
\cite{science:275:1759,pr:d45:1602,prl:82:2484}.
THERA running with the nominal electron beam energy of 250 GeV
(THERA-250) will be only sensitive to mass scales from about 3 to  9 TeV.
With electron beam energy increased to 500 GeV (THERA-500), 
contact interaction mass scale limits improve on average 
by 20-30\%. 
Another improvement by similar factor is observed when going from THERA-500
to THERA-800 option.
Nevertheless, even for high electron beam energies (THERA-500 and THERA-800)
improvement of existing limits will only be possible for selected
models (mainly models coupling to the $u$ quark only).

For contact interaction models conserving parity (Tab.~\ref{tab:afz:cires2}),
current limits from global analysis are, on average, lower than for 
parity violating models.
At the same time THERA sensitivity increases.
Already at THERA-250  mass scale limits can be improved for about
half of the considered models, provided that both $e^{-}p$ and $e^{+}p$  data
are collected.
With increasing electron beam energy, most limits can be significantly
improved. 
From combined $e^{-}p$ and $e^{+}p$  data at THERA-800 limits on
the contact interaction mass scales up to about 18 TeV can be obtained.

For the model of large extra dimensions, THERA will improve existing limits
in any configuration. 
This is because the graviton exchange contribution increases with the
increasing center-of-mass energy.
THERA-800 will be sensitive to the effective Plank scale $M_{S}$ up to
about 2.8 TeV.

In the limit of heavy leptoquark masses ($M_{LQ} \gg \sqrt{s}$)
contact interaction model has been also used to set limits on
the leptoquark mass to the coupling ratio  $M_{LQ}/\lambda_{LQ}$.
Expected 95\% CL exclusion limits, for different leptoquark models
and different THERA running scenarios are presented in Tab.~\ref{tab:afz:lqres}.
Current limits from the global analysis \cite{epj:c17:695} 
are included for comparison.
In most cases existing limits are already above THERA sensitivity, even
in the highest electron energy option.
Limits on  $M_{LQ}/\lambda_{LQ}$ can be only improved for $\tilde{V}_{\circ}$
and  $\tilde{V}_{1/2}$ models.

For leptoquark masses comparable with the available center-of-mass energy,
contact interaction approach has to be modified, as described in
Sect.~\ref{sec:afz:lq}.
Limits on $\lambda_{LQ}$ are calculated as a function of
the leptoquark mass $M_{LQ}$ from the two-dimensional 
probability function ${\cal P}(\lambda_{LQ},M_{LQ})$.
For leptoquark masses smaller than the available center-of-mass energy
($M_{LQ} < \sqrt{s}$),
limits are also set from the measurement of the direct 
leptoquark production.
It turns out both approaches give similar results \cite{epj:c17:695}.
Measurement of the direct leptoquark production process results in better 
limits for low leptoquark masses ($M_{LQ} \ll \sqrt{s}$) and
for leptoquark production involving valence quarks (production of 
F=2 leptoquarks in $e^{-}p$  collisions or F=0 leptoquarks in $e^{+}p$
collisions).
$\frac{d\sigma}{dQ^{2}}$ measurement can results in slightly better 
limits (than expected from the direct production process) 
for leptoquark masses close to the center-of-mass energy and
for leptoquark production from anti-quarks in the proton (F=0 leptoquarks 
in $e^{-}p$ or F=2 in $e^{+}p$).
For leptoquark masses $M_{LQ} < \sqrt{s}$ both kinds of limits are
always calculated and the stronger one is taken.

\begin{figure}[p]
\centerline{\resizebox{\textwidth}{!}{%
  \includegraphics{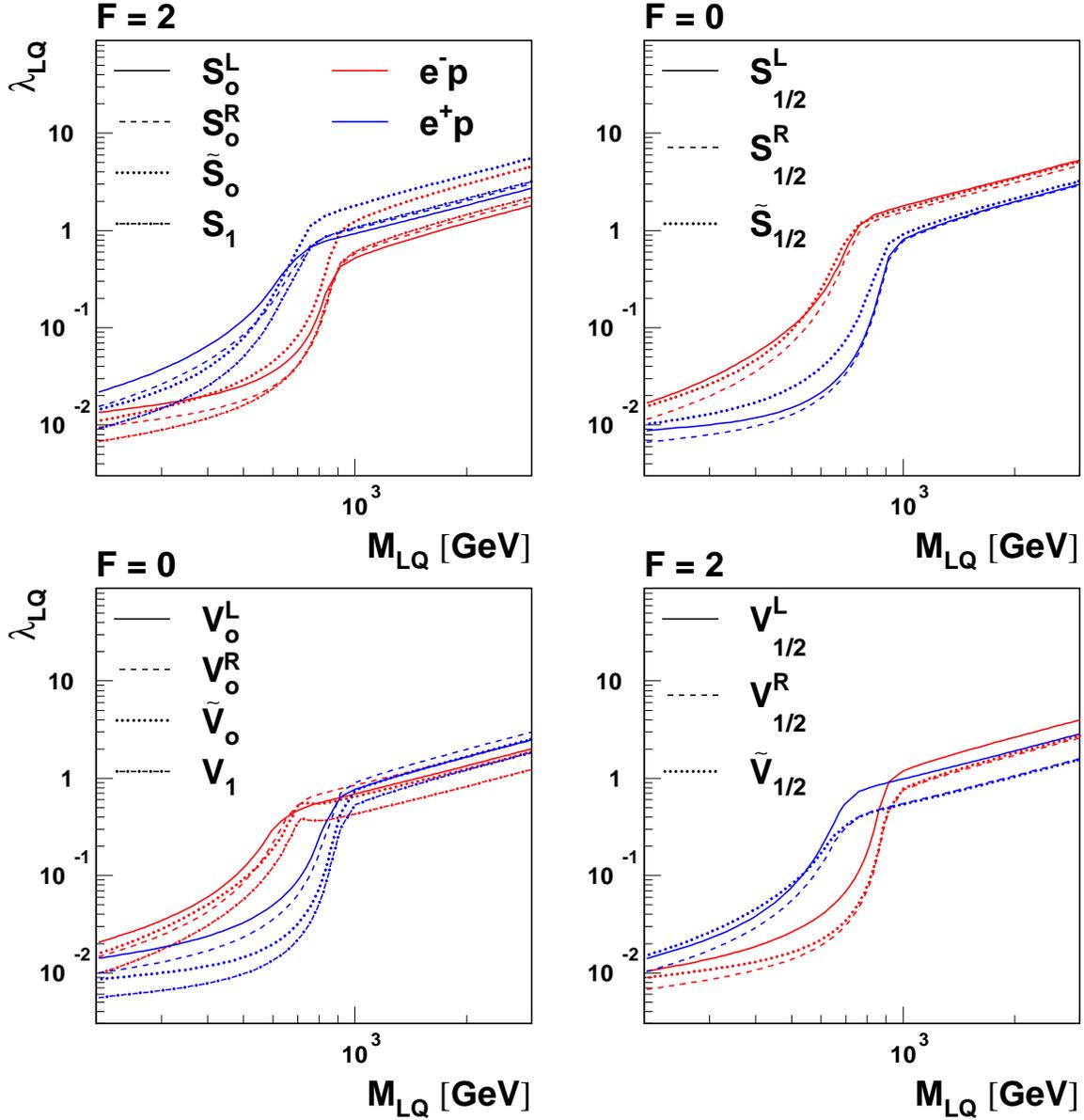}
}}
  \caption{
            Expected 95\% CL exclusion limits in  $(\lambda_{LQ},M_{LQ})$
            space, for different leptoquark  models (as indicated in the plot),
            for 100 $pb^{-1}$ of $e^{-}p$ data (red curves)
            and 100 $pb^{-1}$ of $e^{+}p$ data (blue curves) 
            collected at 250 GeV electron (positron) beam energy
            (THERA-250). Limits based on single leptoquark production 
            and high-$Q^{2}$ NC DIS cross-section  measurements. 
           }
  \label{fig:afz:lq250}
\end{figure}

\begin{figure}[p]
\centerline{\resizebox{\textwidth}{!}{%
  \includegraphics{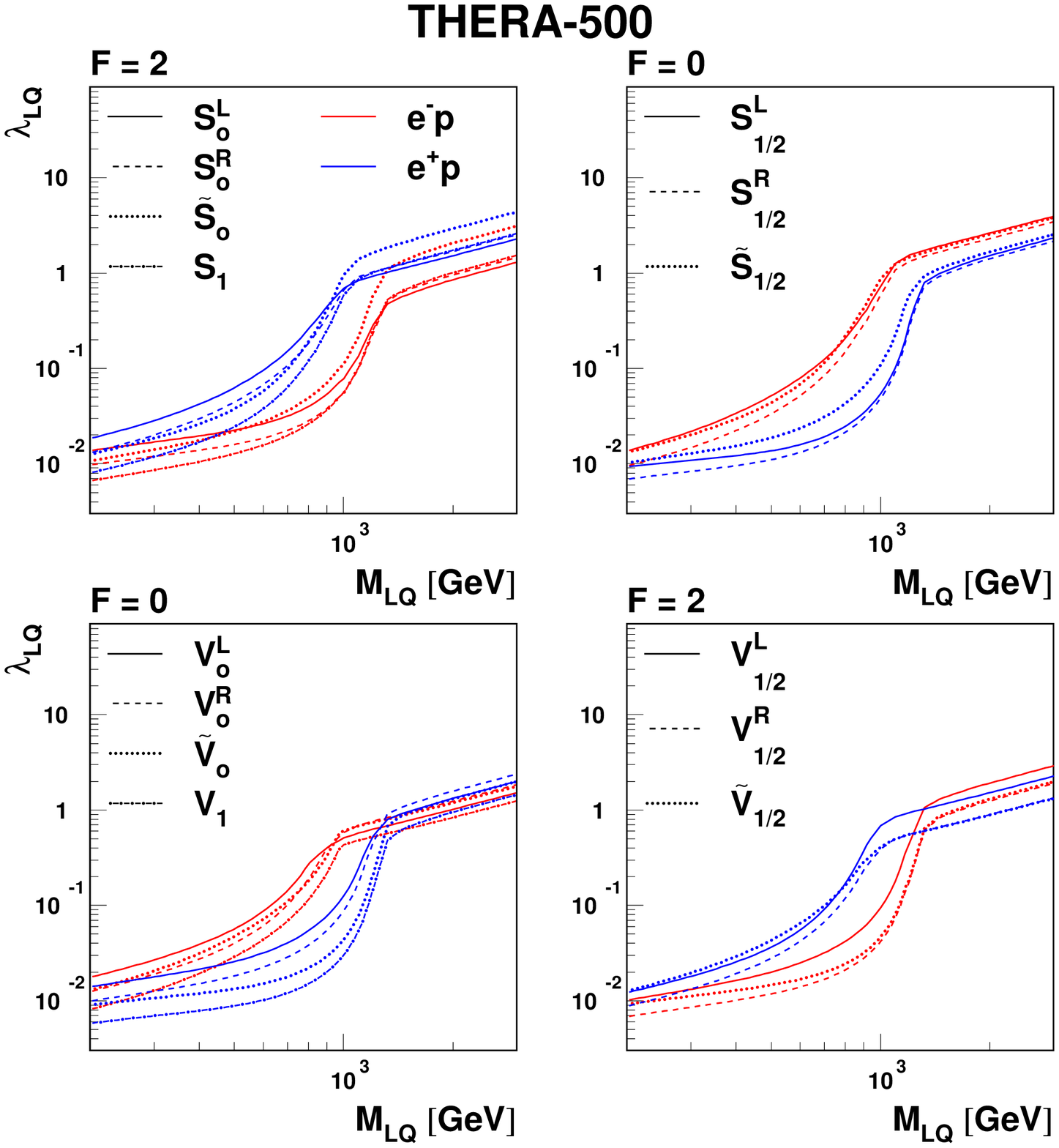}
}}
  \caption{
            Expected 95\% CL exclusion limits in  $(\lambda_{LQ},M_{LQ})$
            space, for different leptoquark  models (as indicated in the plot),
            for 100 $pb^{-1}$ of $e^{-}p$ data (red curves)
            and 100 $pb^{-1}$ of $e^{+}p$ data (blue curves) 
            collected at 500 GeV electron (positron) beam energy
            (THERA-500). Limits based on single leptoquark production 
            and high-$Q^{2}$ NC DIS cross-section  measurements. 
          }
  \label{fig:afz:lq500}
\end{figure}

\begin{figure}[p]
\centerline{\resizebox{\textwidth}{!}{%
  \includegraphics{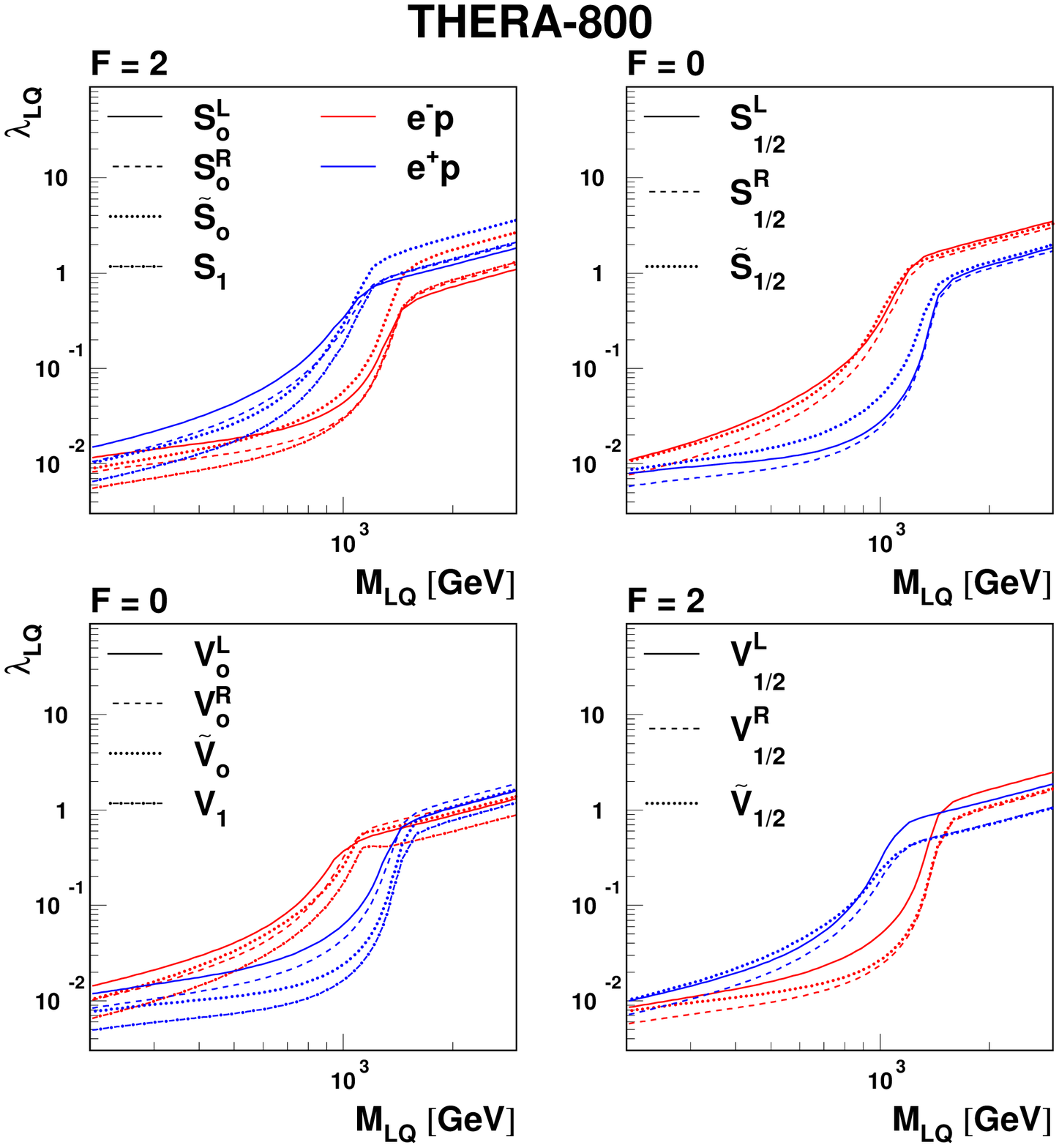}
}}
  \caption{
            Expected 95\% CL exclusion limits in  $(\lambda_{LQ},M_{LQ})$
            space, for different leptoquark  models (as indicated in the plot),
            for 200 $pb^{-1}$ of $e^{-}p$ data (red curves)
            and 200 $pb^{-1}$ of $e^{+}p$ data (blue curves) 
            collected at 800 GeV electron (positron) beam energy
            (THERA-800). Limits based on single leptoquark production 
            and high-$Q^{2}$ NC DIS cross-section  measurements. 
          }
  \label{fig:afz:lq800}
\end{figure}

\begin{figure}[p]
\centerline{\resizebox{\textwidth}{!}{%
  \includegraphics{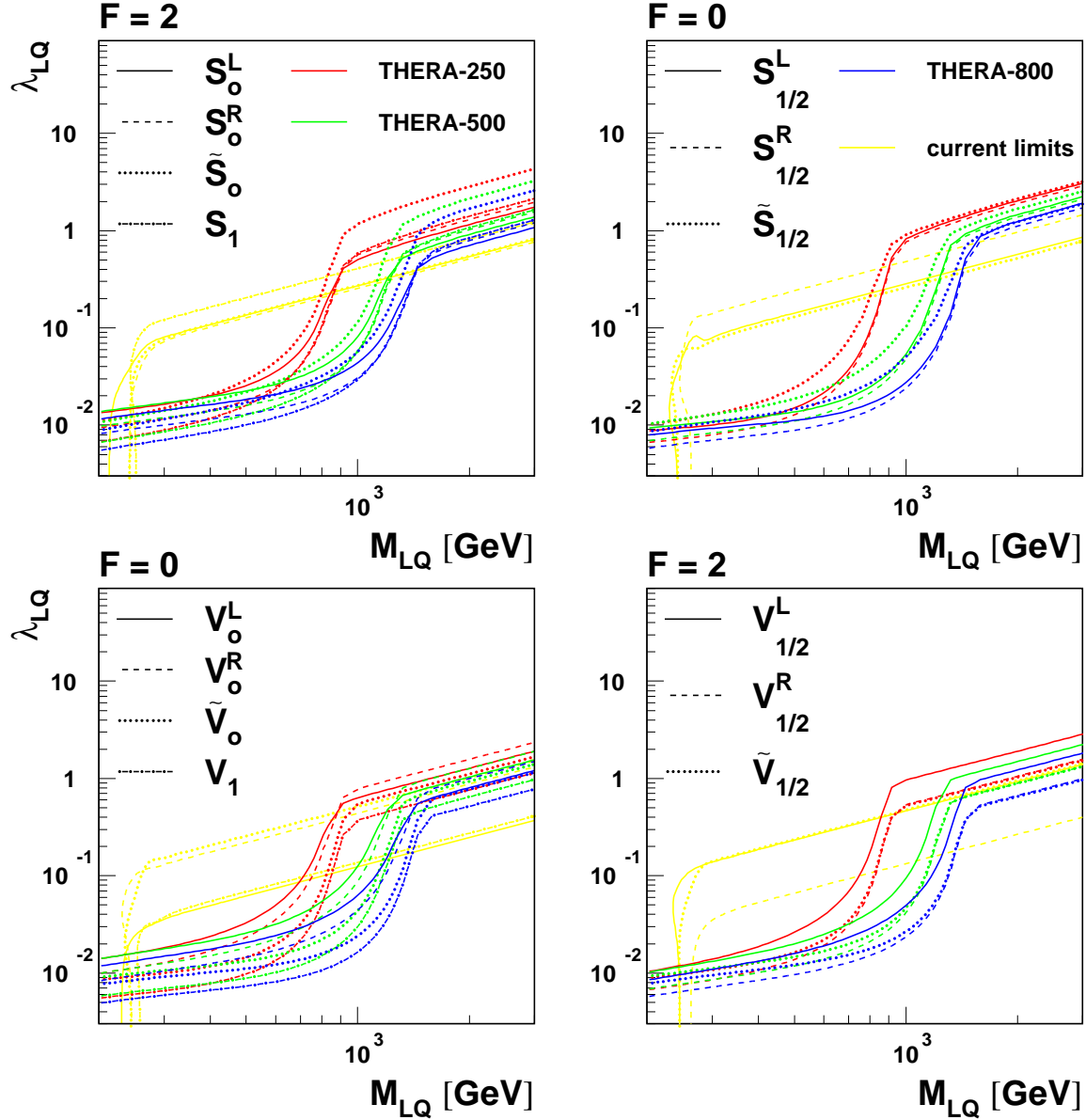}
}}
  \caption{
            Expected 95\% CL exclusion limits in  $(\lambda_{LQ},M_{LQ})$
            space, for different leptoquark  models and for different 
            THERA running scenarios  (as indicated in the plot).
            Limits based on single leptoquark production 
            and high-$Q^{2}$ NC DIS cross-section  measurements
            from the combined $e^{-}p$ and $e^{+}p$ data.
            Indicated in yellow are existing limits from global analysis
            \protect\cite{epj:c17:695}.
          }
  \label{fig:afz:lqene}
\end{figure}

Shown in Fig.~\ref{fig:afz:lq250}, \ref{fig:afz:lq500} and \ref{fig:afz:lq800} 
are expected 95\% CL exclusion limits in  $(\lambda_{LQ},M_{LQ})$, 
for different leptoquark models and THERA running with 250~GeV,
500~GeV and 800~GeV electron (positron) beam respectively.
Limits expected from $e^{-}p$ data and from $e^{+}p$ data are compared.
As expected, better limits on the F=2 leptoquark Yukawa coupling
$\lambda_{LQ}$, for  $M_{LQ} < \sqrt{s}$, are obtained from $e^{-}p$ data, 
whereas $e^{+}p$ data constrain better F=0 leptoquarks.
Differences between limits expected from $e^{-}p$ and  $e^{+}p$ data
are smaller for scalar leptoquarks with $M_{LQ} > \sqrt{s}$.
For high-mass vector leptoquarks it turns out that better limits 
can be obtained for ``wrong'' beam choice ($e^{-}p$ for 
F=0 leptoquarks and $e^{+}p$ for F=2 leptoquarks).
Limits expected from combined $e^{-}p$ and $e^{+}p$ data, 
for different THERA running scenarios, are compared with existing 
limits \cite{epj:c17:695} in Fig.~\ref{fig:afz:lqene}.
In all cases search for single leptoquark production at THERA
significantly improves the existing limits.

\addtolength{\tabcolsep}{1mm}
\begin{table}[t]
 \begin{center}
   \begin{tabular}{lcrrrcrrrcrrr}
      \hline\hline\hline\noalign{\smallskip}
  & Current  & \multicolumn{11}{c}{ Expected 95\% CL exclusion limits on 
$M_{LQ}/\lambda_{LQ}$ [TeV]} \\ \cline{3-13}\noalign{\smallskip}
 Model & limit & 
 \multicolumn{3}{c}{\sc Thera-250} & &  \multicolumn{3}{c}{\sc Thera-500} & &
   \multicolumn{3}{c}{\sc Thera-800} \\
 \cline{3-5} \cline{7-9} \cline{11-13} \noalign{\smallskip}
 & [TeV]
 & $e^{-}p$ & $e^{+}p$ & $e^{\pm}p$  & & $e^{-}p$ & $e^{+}p$ & $e^{\pm}p$ &
 & $e^{-}p$ & $e^{+}p$ & $e^{\pm}p$\\
\noalign{\smallskip}
\hline\hline\hline\noalign{\smallskip}
%
% Results from files:
% tera_250_ele_lq.tex tera_250_pos_lq.tex tera_250_comb_lq.tex tera_500_ele_lq.tex tera_500_pos_lq.tex tera_500_comb_lq.tex tera_800_ele_lq.tex tera_800_pos_lq.tex tera_800_comb_lq.tex
%
$S_{\circ}^{L}     $     & 3.7 &
1.7 & 1.2 & 1.7  & & 2.1 & 1.5 & 2.2  & & 2.5 & 1.8 & 2.7 \\
$S_{\circ}^{R}     $     & 3.9 &
1.5 & 1.1 & 1.5  & & 1.8 & 1.4 & 1.9  & & 2.2 & 1.6 & 2.4 \\
$\tilde{S}_{\circ} $     & 3.6 &
0.7 & 0.6 & 0.7  & & 0.9 & 0.8 & 0.9  & & 1.1 & 0.9 & 1.1 \\
$S_{1/2}^{L}       $     & 3.5 &
0.6 & 1.0 & 1.0  & & 0.8 & 1.2 & 1.2  & & 0.9 & 1.5 & 1.5 \\
$S_{1/2}^{R}       $     & 2.1 &
0.7 & 1.0 & 1.0  & & 0.9 & 1.3 & 1.3  & & 1.0 & 1.6 & 1.6 \\
$\tilde{S}_{1/2}   $     & 3.8 &
0.6 & 1.0 & 1.0  & & 0.8 & 1.2 & 1.2  & & 1.0 & 1.5 & 1.5 \\
$S_{1}             $     & 2.4 &
1.4 & 1.0 & 1.4  & & 1.7 & 1.2 & 1.7  & & 2.1 & 1.5 & 2.1 \\
\noalign{\smallskip}
\hline\noalign{\smallskip}
$V_{\circ}^{L}     $     & 8.1 &
1.6 & 1.4 & 1.8  & & 2.1 & 1.8 & 2.3  & & 2.5 & 2.2 & 2.8 \\
$V_{\circ}^{R}     $     & 2.3 &
1.3 & 1.1 & 1.4  & & 1.7 & 1.4 & 1.8  & & 2.0 & 1.7 & 2.1 \\
$\tilde{V}_{\circ} $     & 1.9 &
1.8 & 1.2 & 1.9  & & 2.1 & 1.6 & 2.4  & & 2.7 & 1.9 & 2.9 \\
$V_{1/2}^{L}       $     & 2.1 &
0.8 & 1.1 & 1.1  & & 1.0 & 1.4 & 1.4  & & 1.2 & 1.7 & 1.7 \\
$V_{1/2}^{R}       $     & 7.5 &
1.2 & 2.0 & 2.0  & & 1.5 & 2.5 & 2.5  & & 1.8 & 3.0 & 3.1 \\
$\tilde{V}_{1/2}   $     & 2.1 &
1.1 & 2.0 & 2.1  & & 1.4 & 2.5 & 2.5  & & 1.6 & 3.1 & 3.1 \\
$V_{1}             $     & 7.3 &
2.6 & 1.5 & 2.8  & & 3.1 & 1.9 & 3.3  & & 3.9 & 2.3 & 4.2 \\
\noalign{\smallskip}
\hline\hline\hline\noalign{\smallskip}
    \end{tabular}
   \end{center}
  \caption{95\% CL exclusion limits on $M_{LQ}/\lambda_{LQ}$ 
         (in the limit of heavy leptoquark masses $M_{LQ} \gg \sqrt{s}$)
         expected from the
         measurement of high-$Q^{2}$ NC DIS cross-sections 
         at  THERA, for different running scenarios, as indicated
         in the table.}
  \label{tab:afz:lqres}
\end{table}
\addtolength{\tabcolsep}{1mm}

\begin{figure}[p]
\centerline{\resizebox{\textwidth}{!}{%
  \includegraphics{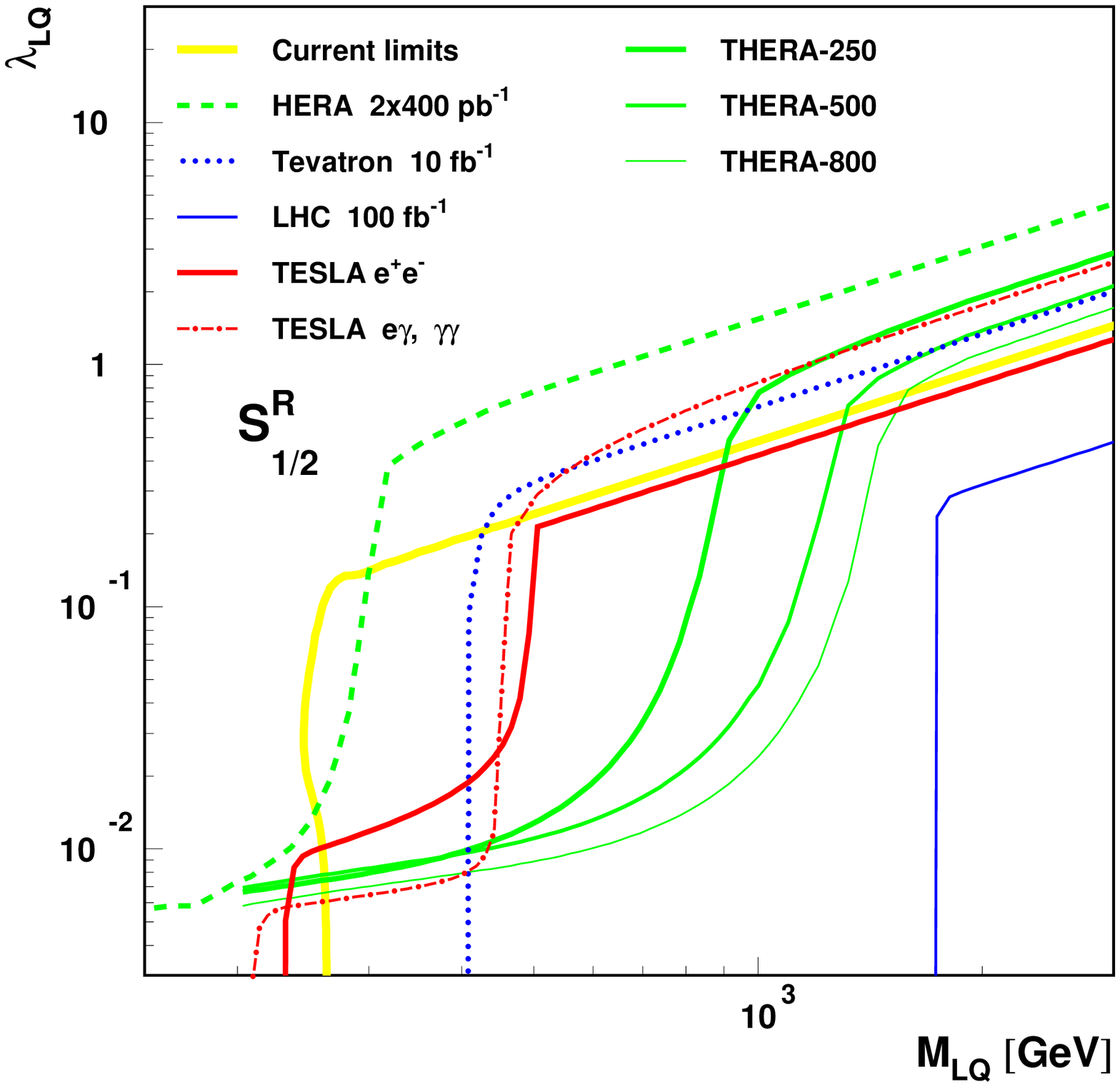}
}}
  \caption{
          Comparison of expected 95\% CL exclusion limits 
            in  $(\lambda_{LQ},M_{LQ})$ for $S_{1/2}^R$ leptoquark  model,
            for different THERA running scenarios and other 
            future experiments, as indicated in the plot.
            Presented limits correspond to 
2$\times$400 $pb^{-1}$ of $e^{\pm}p$ data at HERA ($\sqrt{s}$=318 GeV),
2$\times$100 $pb^{-1}$ or 2$\times$200 $pb^{-1}$ of $e^{\pm}p$ data at THERA 
($\sqrt{s}$=1.0,1.4 and 1.6 TeV),
10 $fb^{-1}$ of $p\bar{p}$ data at the Tevatron ($\sqrt{s}$=2 TeV),
100 $fb^{-1}$ of $pp$ data at the LHC ($\sqrt{s}$=14 TeV)
and 100 $fb^{-1}$ of $e^{+}e^{-}$, $e\gamma$ and $\gamma\gamma$ data 
      at TESLA ($\sqrt{s_{ee}}$=500 GeV).
  Also indicated are 95\% CL exclusion limits from global analysis 
  of existing data\protect\cite{epj:c17:695}.
           }
  \label{fig:afz:lq5}
\end{figure}

\begin{figure}[p]
\centerline{\resizebox{\textwidth}{!}{%
  \includegraphics{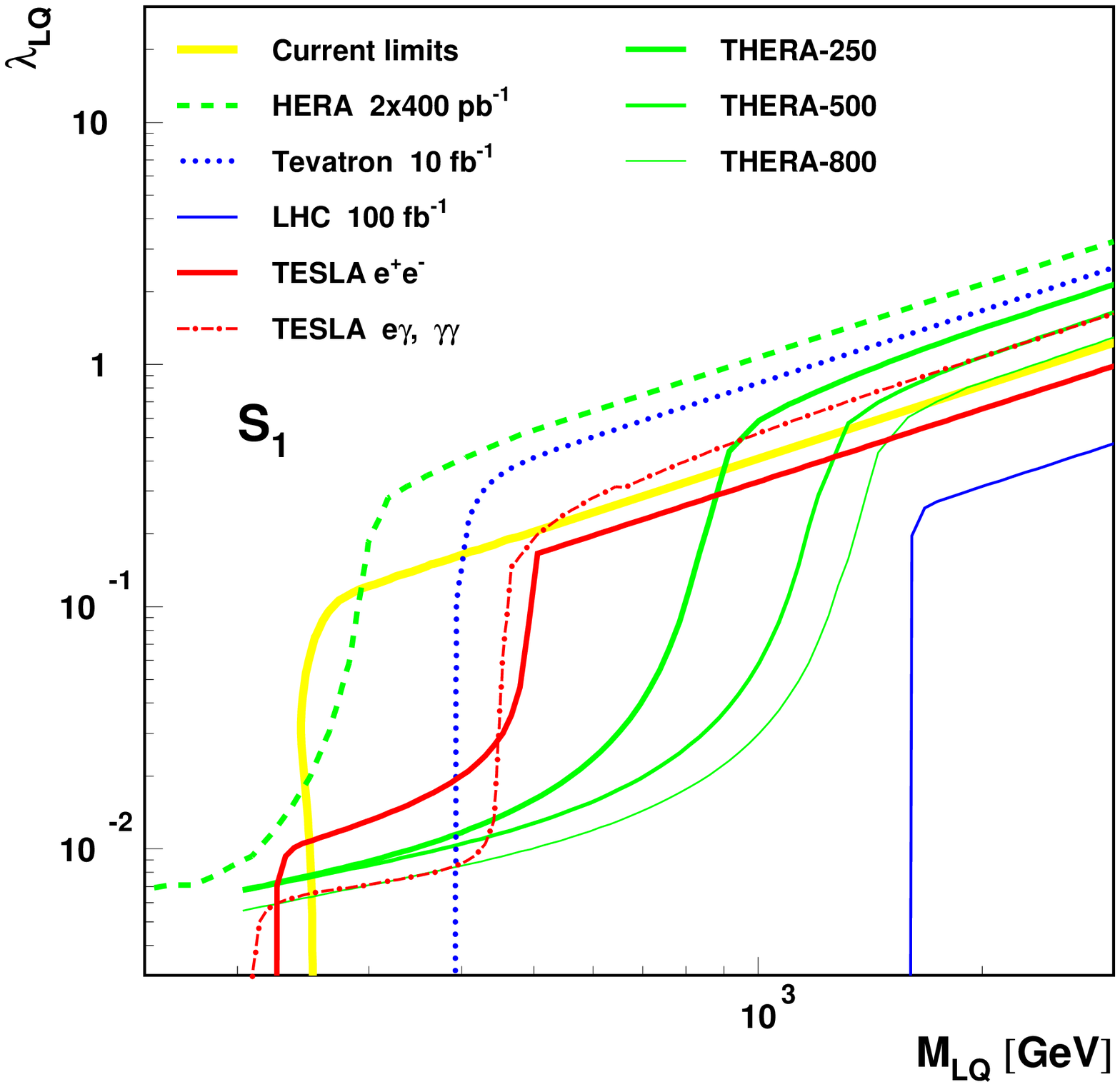}
}}
  \caption{
          Comparison of expected 95\% CL exclusion limits 
            in  $(\lambda_{LQ},M_{LQ})$ for $S_{1}$ leptoquark  model,
            for different THERA running scenarios and other 
            future experiments, as indicated in the plot.
            Presented limits correspond to 
2$\times$400 $pb^{-1}$ of $e^{\pm}p$ data at HERA ($\sqrt{s}$=318 GeV),
2$\times$100 $pb^{-1}$ or 2$\times$200 $pb^{-1}$ of $e^{\pm}p$ data at THERA 
($\sqrt{s}$=1.0,1.4 and 1.6 TeV),
10 $fb^{-1}$ of $p\bar{p}$ data at the Tevatron ($\sqrt{s}$=2 TeV),
100 $fb^{-1}$ of $pp$ data at the LHC ($\sqrt{s}$=14 TeV)
and 100 $fb^{-1}$ of $e^{+}e^{-}$, $e\gamma$ and $\gamma\gamma$ data 
      at TESLA ($\sqrt{s_{ee}}$=500 GeV).
  Also indicated are 95\% CL exclusion limits from global analysis 
  of existing data\protect\cite{epj:c17:695}.
}
  \label{fig:afz:lq7}
\end{figure}

\begin{figure}[p]
\centerline{\resizebox{\textwidth}{!}{%
  \includegraphics{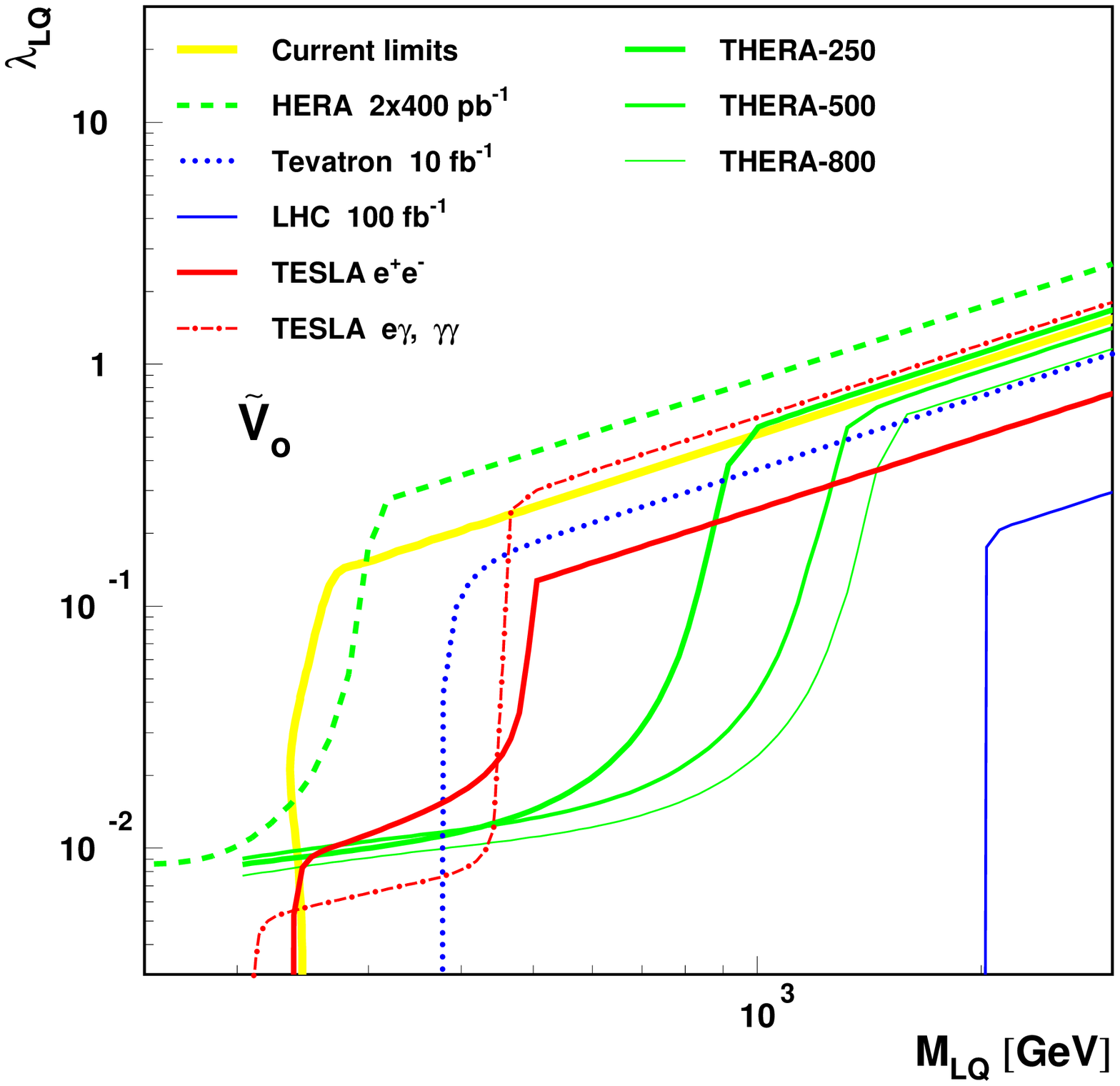}
}}
  \caption{
          Comparison of expected 95\% CL exclusion limits 
        in  $(\lambda_{LQ},M_{LQ})$ for $\tilde{V}_{\circ}$ leptoquark  model,
            for different THERA running scenarios and other 
            future experiments, as indicated in the plot.
            Presented limits correspond to 
2$\times$400 $pb^{-1}$ of $e^{\pm}p$ data at HERA ($\sqrt{s}$=318 GeV),
2$\times$100 $pb^{-1}$ or 2$\times$200 $pb^{-1}$ of $e^{\pm}p$ data at THERA 
($\sqrt{s}$=1.0,1.4 and 1.6 TeV),
10 $fb^{-1}$ of $p\bar{p}$ data at the Tevatron ($\sqrt{s}$=2 TeV),
100 $fb^{-1}$ of $pp$ data at the LHC ($\sqrt{s}$=14 TeV)
and 100 $fb^{-1}$ of $e^{+}e^{-}$, $e\gamma$ and $\gamma\gamma$ data 
      at TESLA ($\sqrt{s_{ee}}$=500 GeV).
  Also indicated are 95\% CL exclusion limits from global analysis 
  of existing data\protect\cite{epj:c17:695}.
}
  \label{fig:afz:lq10}
\end{figure}

\begin{figure}[p]
\centerline{\resizebox{\textwidth}{!}{%
  \includegraphics{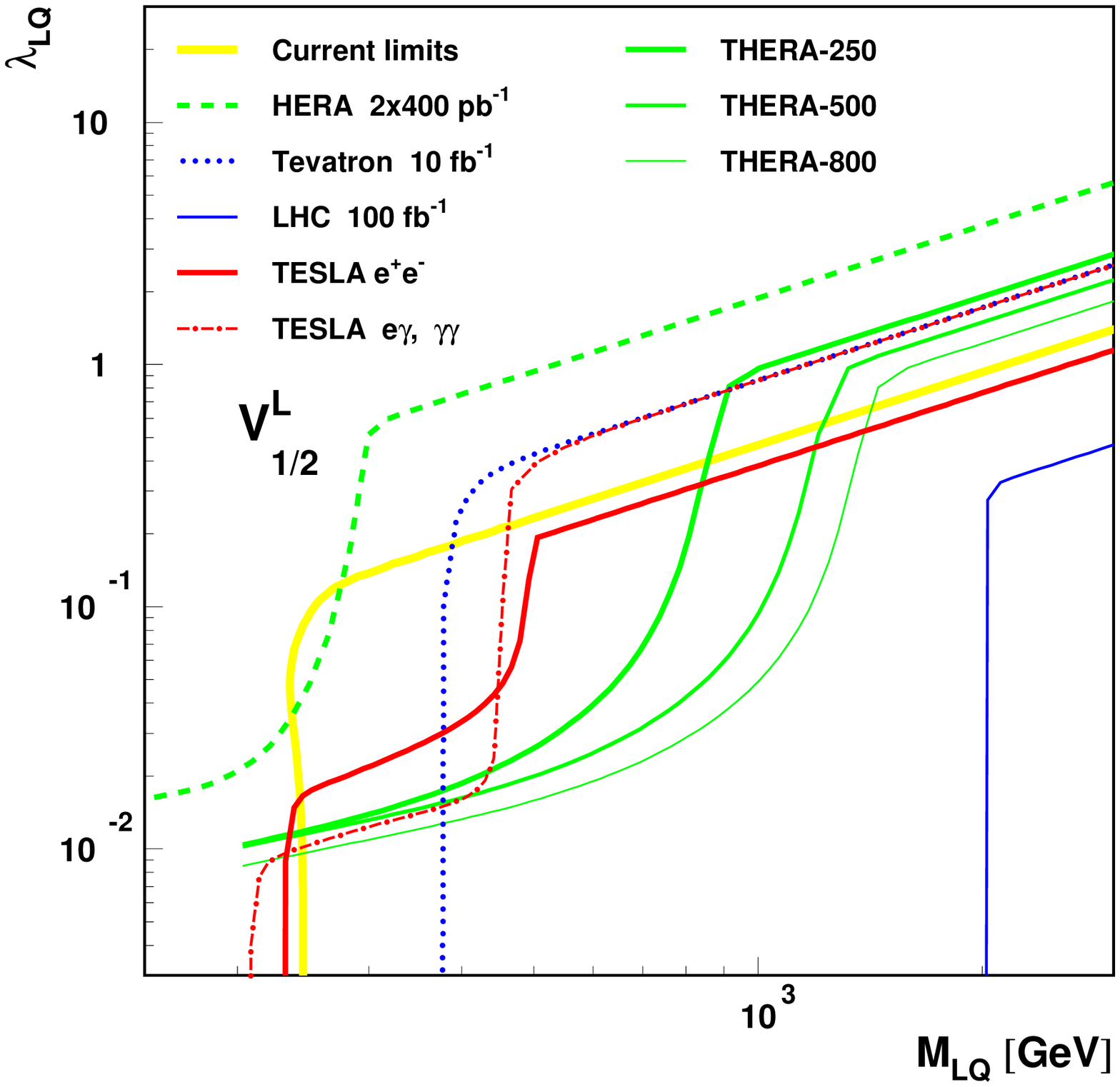}
}}
  \caption{
          Comparison of expected 95\% CL exclusion limits 
            in  $(\lambda_{LQ},M_{LQ})$ for $V_{1/2}^L$ leptoquark  model,
            for different THERA running scenarios and other 
            future experiments, as indicated in the plot.
            Presented limits correspond to 
2$\times$400 $pb^{-1}$ of $e^{\pm}p$ data at HERA ($\sqrt{s}$=318 GeV),
2$\times$100 $pb^{-1}$ or 2$\times$200 $pb^{-1}$ of $e^{\pm}p$ data at THERA 
($\sqrt{s}$=1.0,1.4 and 1.6 TeV),
10 $fb^{-1}$ of $p\bar{p}$ data at the Tevatron ($\sqrt{s}$=2 TeV),
100 $fb^{-1}$ of $pp$ data at the LHC ($\sqrt{s}$=14 TeV)
and 100 $fb^{-1}$ of $e^{+}e^{-}$, $e\gamma$ and $\gamma\gamma$ data 
      at TESLA ($\sqrt{s_{ee}}$=500 GeV).
  Also indicated are 95\% CL exclusion limits from global analysis 
  of existing data\protect\cite{epj:c17:695}.
}
  \label{fig:afz:lq11}
\end{figure}

In Fig.~\ref{fig:afz:lq5}, \ref{fig:afz:lq7}, \ref{fig:afz:lq10} and 
\ref{fig:afz:lq11},  limits on the leptoquark Yukawa coupling $\lambda_{LQ}$
and mass $M_{LQ}$ expected from THERA are compared with existing limits
and limits expected from other future experiments \cite{hep-ph-0006335},
for $S_{1/2}^R$, $S_1$, $\tilde{V}_{\circ}$
and $V_{1/2}^L$ leptoquark models respectively.\footnote{
Selected models were shown to describe the existing experimental
data better than the Standard Model \cite{epj:c17:695}.}
Leptoquarks with masses up to about 2.0 TeV can be searched for
at LHC, independently of $\lambda_{LQ}$.
THERA will not be able to improve any limits if LHC excludes
leptoquark masses below 1.6 TeV.
However, if any leptoquark type state is discovered at LHC,
THERA will be the best place to study its properties,
covering the widest range in $(\lambda_{LQ},M_{LQ})$ space.
Leptoquark mass, spin, fermion number and  branching fraction
(assuming leptoquark decays into $\nu + jet$ are reconstructed)
can be  determined.
Yukawa coupling can be precisely measured down to the very small 
coupling values of the order of $\lambda_{LQ} \sim 10^{-2}$,
not accessible at LHC.

%
%---------------------------------------------------------------------------
%
\section{Summary}

The sensitivity of THERA to different contact interaction models
has been studied in detail.
For models conserving parity, scales up to about 18 TeV can be 
probed at THERA, extending considerably beyond the existing bounds.
Significant improvement of existing limits is also expected for models
with large extra dimensions.
Effective Plank mass scales up to about 2.8 TeV can be probed.
THERA will be the best machine to study leptoquark properties,
for leptoquark masses up to about 1 TeV.
It will be sensitive to the leptoquark Yukawa couplings down to
$\lambda_{LQ} \sim 10^{-2}$.

%
%---------------------------------------------------------------------------
%

% \clearpage

\bibliography{afz_all}

%---------------------------------------------------------------------------

\end{document}